\documentclass[aps,prl,twocolumn,amsmath,amssymb]{revtex4-1}
\usepackage{graphicx}
\usepackage{dcolumn}
\usepackage{bm}
\usepackage{epsfig}

\begin{document}

\title{THz emission from intrinsic Josephson junctions at zero magnetic field via breather auto-oscillations.}

\author{V. M. Krasnov}
\affiliation{ Department of Physics, AlbaNova University Center,
Stockholm University, SE-10691 Stockholm, Sweden }


\begin{abstract}

I propose a new mechanism of intense high-frequency
electromagnetic wave generation by spatially uniform stacked
Josephson junctions at zero magnetic field.
The ac-Josephson effect converts the dc-bias voltage
into ac-supercurrent, however, in the absence of spatial variation of Josephson phase difference, does not
provide dc-to-ac power conversion, needed for emission of electromagnetic waves. Here I demonstrate that at geometrical resonance conditions,
the spatial homogeneity of phase can be spontaneously broken by
appearance of breathers (bound fluxon-antifluxon pairs), facilitating effective dc-to-ac power conversion.
The proposed mechanism explains all major features of recently observed THz
radiation from large area Bi$_2$Sr$_2$CaCu$_2$O$_{8+x}$ mesa
structures.
\end{abstract}

\maketitle

Auto-oscillations are self-sustaining oscillations in nonlinear
systems with frequency and often amplitude independent of
driving force. Oscillatory behavior in plasma \cite{Plasma},
operation of a clock and multivibrators, voice and sound of
musical instruments, all are examples of
auto-oscillations. Typically, periodic auto-oscillations are
excited by a constant force. The frequency is
determined either by an internal resonance (e.g. a cavity mode),
or a characteristic relaxation time. Auto-oscillations are widely used for generation of
microwaves, e.g. in Gunn diodes, magnetrons and klystrons.

Recently a significant THz emission has been reported at zero
magnetic field from large area
$\text{Bi}_{2}\text{Sr}_{2}\text{Ca}\text{Cu}_{2}\text{O}_{8+x}$
(Bi-2212) mesa structures \cite{Ozyuzer,Wang}, which represent
natural stacks of atomic-scale intrinsic Josephson junctions
(IJJs). 
Although the emission occurs at
geometrical (Fiske) resonance frequencies, association with conventional Fiske steps \cite{SvenFiske}
is problematic because their amplitude is zero at $H=0$. Furthermore, significant
emission power 
would require a large quality factor $Q\gg 1$ \cite{FiskeTheory}.
Since $Q$ of Fiske steps is inversely proportional to the mesa size \cite{SvenFiske}, it is not clear how it could be
sufficiently large for large mesas at large operation temperatures and quasiparticle
damping due to large self-heating. Therefore, the observed
radiation is quite puzzling and remains a mater of intense discussion
\cite{FiskeTheory,Hu,KoshelevNonUn,Koshelev2pi,HotSpot,Tachiki2009,TTachiki2010,Klemm2010}.

In this letter I propose a new mechanism of emission from a
spatially uniform stack of Josephson junctions at zero applied
field. It is shown that at geometrical resonance conditions the
homogeneous state becomes unstable with
respect to formation of a breather lattice in the stack. Breathers
are bound fluxon-antifluxon pairs \cite{McLaughlinScott}. They
effectively couple the dc-bias to the ac-Josephson oscillations
and makes resonances self-sustainable, once ignited. It is argued that the
discovered breather auto-oscillations explain experimental
features of the zero-field emission from large Bi-2212 mesas.

Radiation from large IJJs is following the
ac-Josephson relation \cite{Ozyuzer,Wang}.
Although the ac-Josephson effect converts the dc-bias voltage into
the oscillating supercurrent, this does not ensure radiation from
junctions because the supercurrent is non-dissipative and can
not by itself pump energy from the dc-power supply into radiation.
Such a dc-to-ac power conversion can be achieved via the Lorentz
force
\begin{equation}\label{Florentz}
F_L=s I \times B,
\end{equation}
(per junction), where $I$ is the dc-bias current and $s$ is the stacking
periodicity ($s \simeq 1.5$~nm for IJJs). Therefore, emission by
means of the ac-Josephson effect requires {\it finite} $B_i(x)
\neq 0$, which is connected with the finite Josephson phase
gradients $\nabla \varphi_i \neq 0$ ($i$ is the junction index).

Structural inhomogeneity may lead to some coupling
of the dc-bias to the ac-resonance field at $H=0$
\cite{ZeroFiske,KoshelevNonUn}. The inhomogeneity can be caused by
variation of the critical current density, or nonuniform bias current distribution \cite{NonUn}. In large
Bi-2212 mesas the inhomogeneity can also be induced by
uneven self-heating at large bias \cite{HotSpot} and by
defects in Bi-2212 single crystals. Associated
phase gradients can be attributed to self-field
caused by nonuniform current flow. This may lead
to appearance zero-field Fiske steps \cite{ZeroFiske}. However, the observed emission is hardly
explained by zero-field Fiske steps: large flux
quantization field in atomic scale IJJs would require very large inhomogeneity \cite{KoshelevNonUn}.
Furthermore, such inhomogeneity
could hinder mutual synchronization of IJJs
\cite{SvenFiske} and suppress collective Fiske resonances,
required for coherent amplification of the emission power.

Fluxons create a large phase gradient $\simeq \pi/\lambda_J$ ($\lambda_J$ is the Josephson penetration depth) and
thus provide an effective dc-to-ac-power conversion. Usually fluxons are
introduced by applying in-plane
magnetic field. In long Josephson junctions, $L \gg \lambda_J$, they can be trapped at $H=0$. Emission scenarios, involving
quasistatic semi-fluxons/antifluxons \cite{Hu}, or fluxons/antifluxons
\cite{Koshelev2pi} in mesas, were proposed. For a single junction
static fluxon-antifluxon pairs are unstable, because they tend to
collapse and annihilate. But for stacked junction fluxon-antifluxon pairs in neighbour junctions are stable
\cite{KrasnovComp,KleinerAntiFlux}. Fluxon-antifluxon
modes at $H=0$ has been clearly observed in large Bi-2212 mesas,
as they cause multiple valued critical current \cite{KrasnovComp}.
Fluxons experience the Lorentz force, which facilitates efficient pumping
of dc-power into kinetic energy of fluxons. Upon
collision and annihilation of fluxons at junction edges, radiation pulses are produced. The corresponding flux-flow emission is well
studied for single junctions \cite{Koshelets}. Significant flux-flow emission from
stacked junctions requires both stabilization of the
square fluxon lattice and high quality geometrical
resonances \cite{FiskeTheory}, because the emission power is $P_{rad} \propto Q^2$ and the
linewidth $\propto Q^{-1}$. For IJJs, this can only be achieved at
strong fields $H>\Phi_0/\lambda_J s \simeq 2$T, and for
small Bi-2212 mesas \cite{Katterwe,SvenFiske}. None of those
requirements is realized in case of zero-field
emission from large mesas.

Several non-Josephson emission mechanisms at $H=0$ are also known.
Stable fluxon-antifluxon pairs in stacked junctions can be unbound
by a large enough transport current and may start shuttling in the
stack, leading to appearance of zero-field steps (ZFS) in $I-V$
curves \cite{KleinerAntiFlux}. Some emission of
electromagnetic waves does occur at ZFS \cite{PedersenZFS,Chan},
however, it occurs at subharmonics of the Josephson
frequency \cite{PedersenZFS} and the emission power is small
because the fluxon is not annihilated upon collision with the
edge, but is reflected as an antifluxon. Therefore, unlike in the
flux-flow case, only a minor part of fluxons energy can be
radiated. Moreover, emission power at ZFS can not be large because the fluxon will not be reflected if it will loose
significant part of its energy upon the collision with the edge \cite{McLaughlinScott}.
In stacks, ZFS can be accompanied by non-Josephson Cherenkov radiation
\cite{Modes,Cherenkov} due to partial decomposition of
fluxons into travelling plasma waves \cite{Modes,Fluxon}.
Also, recombination of injected nonequilibrium quasiparticles leads to generation of bosons \cite{Iguchi,Cascade}.
The nonequilibrium emission is direct (does not involve
the ac-Josephson effect) and can provide very high efficacy
\cite{NoneqPRL2009}. It is most effective at $H=0$ because it
benefits from the sharp gap singularity in the quasiparticle density of
states.

\begin{figure}[t]
\begin{center}
\includegraphics[width=0.8\linewidth]{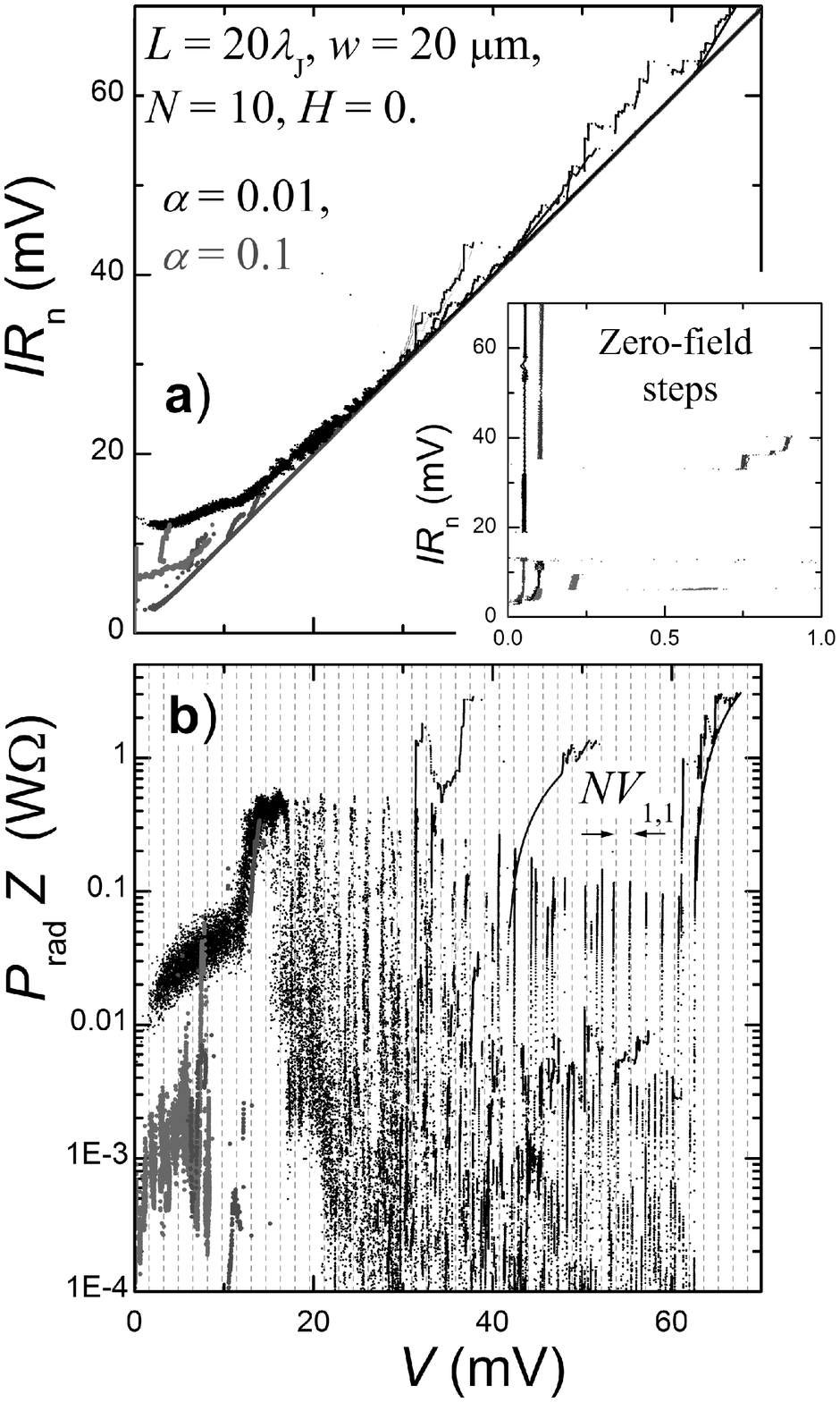}
\end{center} \label{Fig1}
\caption{(Color online). (a) Normalized $I$-$V$ characteristics for long, uniform, underdamped
stack Josephson junctions at zero applied field. Appearance of pronounced steps, caused by breather
auto-oscillations, is seen. Different colors correspond
to simulations with different initial conditions with certain amount of trapped fluxons/antifluxons. Inset shows zero-field
steps at low voltages. (b) Normalized radiative power along the same $I$-$V$'s.
Pronounced maxima occur at collective in-phase cavity resonances,
indicated by grid lines.}
\end{figure}

Analysis of emission from stacked Josephson junctions requires
implementation of proper radiative boundary conditions into the
coupled sine-Gordon equation. In the numerical simulations,
presented below, non-local radiative boundary
conditions, derived in Ref. \cite{FiskeTheory}, are employed. The emission is
facilitated by the finite radiation impedance $Z$. Simulations are
made for a stack of $N=10$ identical, uniform junctions with
parameters typical for optimally doped Bi-2212 IJJs
\cite{Katterwe,SvenFiske}.
The damping parameter 
was varied from strongly underdamped $\alpha =0.01$ to overdamped
$\alpha=1$.
Note that IJJs exhibit hysteresis in $I$-$V$
characteristics, i.e., remain underdamped, almost up to $T_c$
\cite{CollapsePRB}. The emission power is scaling
with the width of the stack $w$ and is normalized to $w=20\mu$m.

\begin{figure}[t]
\begin{center}
\includegraphics[width=0.95\linewidth]{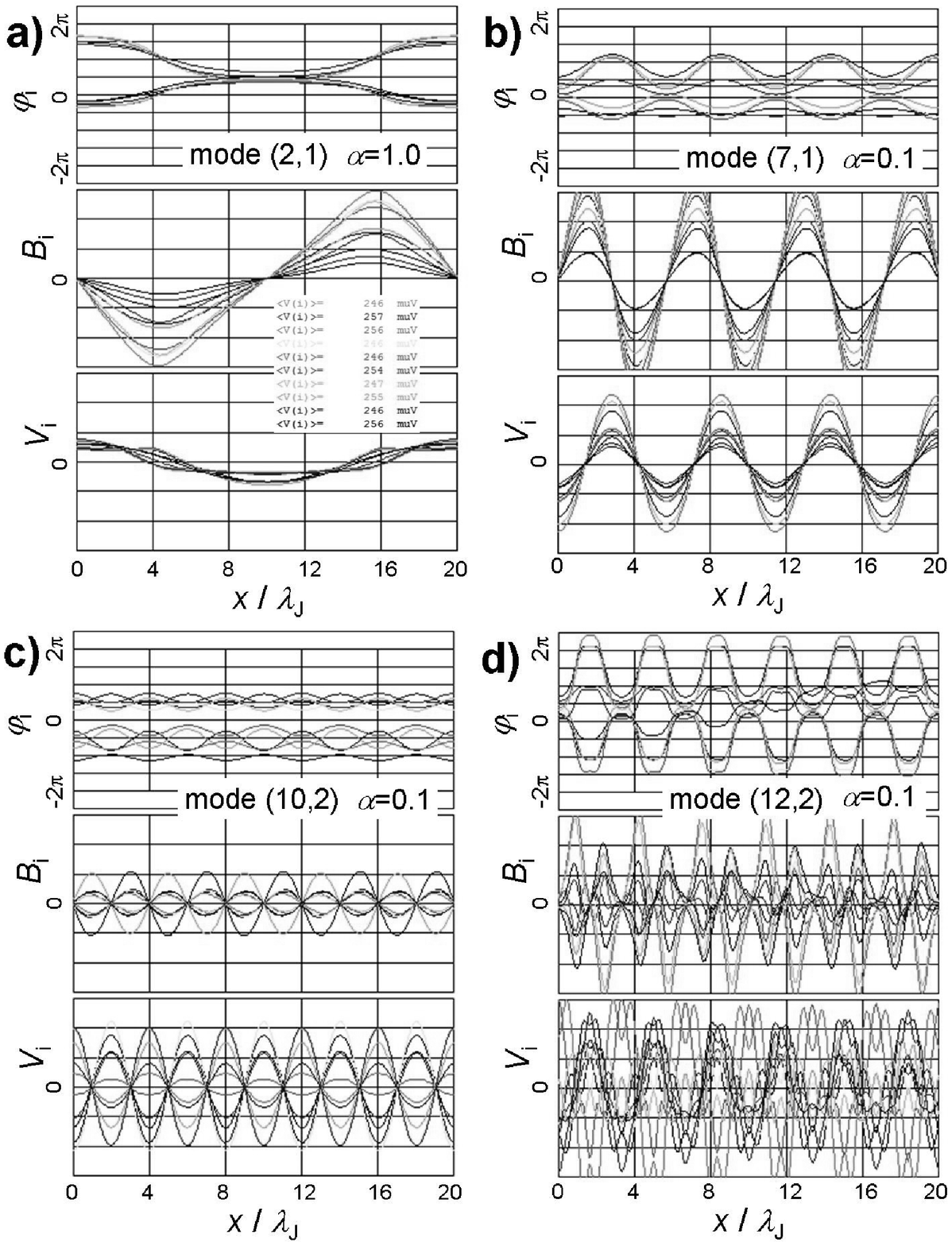}
\end{center}\label{Fig2}
\caption{(Color online). Snapshots of phase (top panels), magnetic
induction (middle panels) and ac-component of voltage (bottom
panels) at four different breather resonances for $L=20\lambda_J$ overdamped $\alpha=1$ (a) and underdamped $\alpha=0.1$ (b-d) stacks.
It is seen that electromagnetic field forms standing-wave patterns in the stack. Resonant cavity modes $(m,n)$ are indicated in the figures. }
\end{figure}

Figure 1 (a) shows $I$-$V$ characteristics for stacks with the length $L=20\lambda_J$ for $\alpha =0.01$ (black,
gray and blue) and $0.1$ (red, magenta). Different
colors represent simulations with different initial conditions, corresponding to certain amount of trapped fluxons and antifluxons.
At low bias this leads to appearance of ZFS
\cite{KleinerAntiFlux}, shown in the inset. At larger bias,
distinct resonances appear, resembling Fiske steps.

Figure 1 (b) shows the radiation power $P_{rad}$ (from one edge), normalized by the radiation impedance $Z$. All
presented simulations are made for very large $Z$, so that radiative losses are much smaller than
internal resistive losses. Under those circumstances, the
product $P_{rad}Z$ is constant \cite{FiskeTheory}. From comparison of Figs. 1 (a) and (b)
it is seen that some steps are accompanied by strong emission,
others to the contrary, correspond to emission minima. Distinct
emission maxima occur at voltages of collective in-phase Fiske
steps, shown by the grid in Fig. 1 (b).

\begin{figure}[t]
\begin{center}
\includegraphics[width=0.7\linewidth]{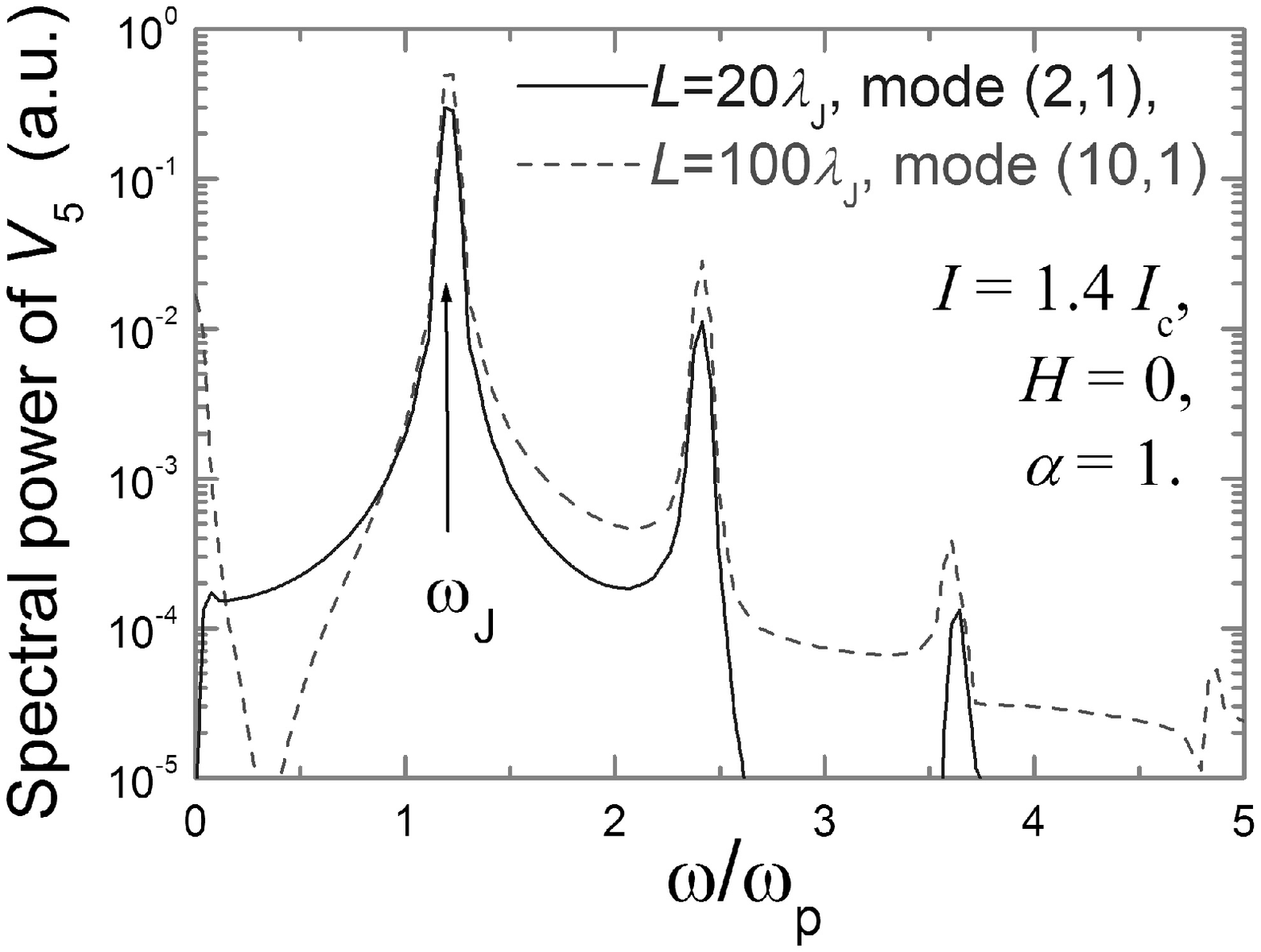}
\end{center}\label{Fig3}
\caption{(Color online). Spectra of voltage oscillations in the middle junction
for overdamped stacks, $\alpha=1$, with $L=20\lambda_J$ and $100\lambda_J$. Stacks are biased at the same dc-voltage,
corresponding to in-phase cavity modes (2,1) and (10,1), respectively.
Spectra are similar and have narrow maxima at the Josephson frequency $\omega_J$.
Note that amplitudes and linewidths of the two resonances are similar, despite different $L$. }
\end{figure}

Figure 2 shows snapshots of spatial distributions of phases $\varphi_i$ (top panels), magnetic inductions
$B_i$ (middle panels, $\simeq 0.184$Gs per division) and ac-components of voltages $V_i$
in each junction (bottom panels, $\simeq 209 \mu$V per division) at for different resonances
for (a) overdamped $\alpha = 1$ and (b-d) underdamped $\alpha = 0.1$ stacks with $L=20\lambda_J$. The
junction color code is represented by the dc-voltage
annotation in Fig. 2 (a). It is seen that electromagnetic
field forms two-dimensional standing wave patterns in the stack,
similar to Fiske resonances \cite{SvenFiske,KleinerModes,FiskeTheory}. Resonant modes $(m,n)$
are characterized by the wave numbers $k_{ab}=m\pi/L$ in-plane and
$k_c=n\pi/Ns$ in the $c$-axis direction \cite{KleinerModes}.

From Fig. 2 it is seen that there is no increment of phase in the
junctions $\varphi_i(x=0)=\varphi_i(L)$, which means that the net
magnetic flux in each junction is zero. The observed modulation of phase
is, therefore, caused by a spontaneous formation of ordered
breather lattice, consisting of similar breather
chains in each junction.
Breather chains couple to the dc-power supply via the Lorentz
force and effectively pump energy into ac-Josephson oscillations.
This makes breather resonances self-sustaining, once ignited. The
frequency of oscillations is adopted to the nearest cavity mode in
the stack. Such behavior is typical for auto-oscillation
phenomena, as mentioned in the introduction. Therefore, I refer
the discovered new type of zero-field resonances to as breather
auto-oscillations.

The breather amplitude is not quantized and the phase variation
can acquire any value in the range $-2\pi <\Delta \varphi_i <
2\pi$. Fig. 2 (a) represents a special case, when the phase
amplitude at a certain time is close to $\pm\pi$, similar to the
case discussed in Ref. \cite{Hu}. Fig. 2 (d) shows another special
case when fluxons and antifluxons in the breather are well
separated and the amplitude $\Delta \varphi_i \simeq \pm
2\pi$, similar to the case considered in
Ref. \cite{Koshelev2pi}. However, even in those cases, it should be
realized that $\Delta \varphi_i$ are not constant but are
oscillating in time. In general the amplitude $\Delta \varphi_i$
can be arbitrary, as shown in Figs. 2 (b,c).

Figure 3 shows the spectra of voltage oscillations in the middle
junction $i=5$ at the edge of the stack $x=0$. Frequency is normalized by the Josephson plasma frequency $\omega_p$. Results are shown
for two overdamped stacks $\alpha=1$ with different lengths
$L=20\lambda_J$ (solid line) and $L=100\lambda_J$ (dashed line),
biased at the same relative current $I/I_c=1.4$ and having the same voltage.
For $L=20\lambda_J$ it corresponds to the in-phase
resonance (2,1), shown in Fig. 2 (a), and for $L=100\lambda_J$ to the in-phase mode (10,1), with the same spatial separation between nodes.
It is seen that in both cases the radiation spectra are similar and consist of a sharp maximum at the primary Josephson frequency, marked by the
upward arrow, and several small harmonics.

Simulations in Fig. 3 demonstrate that breather auto-oscillations are not hampered even in the overdamped case $\alpha = 1$ and in
very long stacks. Furthermore, neither the amplitude, nor the linewidth of oscillations is deteriorated in the longer stack. This implies that
quality factors of the two resonances are similar. Quality factors of geometrical resonances were considered in Ref. \cite{SvenFiske}:
for the mode $(m,n)$, $Q_{m,n}=m\pi(c_n/c_0)(\lambda/L)/\alpha$, where $c_n$ is the
velocity of mode $n$ and $c_0$ is the Swihart velocity of a single
junction. The in-phase mode $n=1$ is the most important for
achieving coherent high power emission. Using the approximate
expression for $c_1$ \cite{SvenFiske}, we obtain
\begin{equation}\label{Qmn}
Q_{m,1} \simeq \frac{m\sqrt{2}(N+1)}{(L/\lambda_J)\alpha},
\end{equation}
valid for $N < \pi\lambda_{ab}/s \simeq 400$, where $\lambda_{ab} \simeq 200$nm is the London penetration depth.
It is seen that modes with the same $L/m$ from Fig. 3 indeed have the same $Q\simeq \sqrt{2} > 1$, despite junctions are long and overdamped.
The quality factor of the junction, $Q_0 \equiv Q(\omega=\omega_p)= 1/\alpha$, should not be confused with the quality factor of geometrical resonances
$Q_{m,n}=(\omega_{m,n}/\omega_p)Q_0$.


In conclusion, a new mechanism of zero-field radiation from
stacked Josephson junctions is proposed, via spontaneous breather
auto-oscillations at geometrical resonance conditions \cite{Supplem}.
It explains all major experimental features of THz emission from
large Bi-2212 mesas \cite{Ozyuzer,Wang}:

i) Breather auto-oscillations lead to powerful coherent radiation
from uniform mesas at zero magnetic
field, because breather can effectively
couple ac-Josephson oscillations to dc-power supply via the
Lorentz force.

ii) They excite conventional cavity modes
in the stack with the frequency following
the ac-Josephson relation and inversely proportional to the
junction length.

iii) Coherent flux-flow emission requires stabilization of the rectangular (in-phase) fluxon lattice,
which is counteracted by fluxon-fluxon repulsion \cite{Katterwe}. Establishment of in-phase breather oscillations
is much easier because of much smaller breather-breather repulsion.

iv) Unlike conventional Fiske resonances, which have only one strong mode for a given field (the velocity matching mode with two nodes per fluxon) \cite{SvenFiske},
the wavelength of breather chain can be arbitrary. This flexibility provides an important advantage with respect to Fiske resonances,
because it allows maximum amplitude for any geometrical resonance at $H=0$.

v) The flexibility of breather resonances ensures that some, high enough, in-phase resonances would have large quality factor, irrespective of
the damping parameter or the junctions length. This allows intense coherent emission of electromagnetic waves with narrow linewidth
even in very large mesas and at elevated temperatures.

vi) Emission due to breather auto-oscillations is
suppressed by small in-plane magnetic field.
Numerical simulations (not shown) demonstrated that
with increasing field, breather resonances are gradually substituted by conventional Fiske
resonances, which have smaller amplitude at low fields.

vii) Breather resonances depend on
initial conditions, which resembles metastability
of emission from Bi-2212 mesas. Essentially breather auto-oscillations need to
be ignited to become self-sustainable. Here they were ignited with the help of
trapped fluxons/antifluxons. In experimental situation it is likely that the ignition is
facilitated by the non-equilibrium mechanism
\cite{Cascade,NoneqPRL2009,Iguchi}, which is most intense at large injection current densities just beneath electrodes, as
seen in laser microscopy \cite{Wang}.

viii) From Eq. (2) it follows that coherent emission is improved in mesas with larger number of IJJs, up to $N\simeq 400$, which is
the optimal number for achieving large emission power, without too large self-heating.

\end{document}